\documentclass[a4paper,11pt]{article}
\usepackage{latexsym}
\usepackage{rotating}
\author{W X Ma\footnote{Email: mawx@cityu.edu.hk}\\
Department of Mathematics, City University of Hong Kong,\\
Kowloon, Hong Kong, China}
\title{A Class of Coupled KdV systems and Their Bi-Hamiltonian Formulations}
\date{\nonumber}

\begin{document}

\setlength{\parskip}{5pt plus 2pt minus 1 pt}
  \setlength{\textwidth}{150.5mm}
  \setlength{\textheight}{230mm}
  \setlength{\oddsidemargin}{0mm}
  \setlength{\topmargin}{-25mm}    
\setlength{\baselineskip}{18pt}
\date{\nonumber}



\centerline{{\Large 
A Class of Coupled KdV Systems and }}
\centerline{{\Large 
Their Bi-Hamiltonian Formulation}}

\vskip 4mm

\centerline{{\large Wen-Xiu Ma}\footnote{Email: mawx@cityu.edu.hk}}
\centerline{{\large Department of Mathematics, City University of Hong Kong,}}
\centerline{{\large Kowloon, Hong Kong}}

\newtheorem{thm}{Theorem}[section]
\newtheorem{Le}{Lemma}[section]
\newtheorem{defi}{Definition}[section]
\newcommand{\R}{\mbox{\rm I \hspace{-0.9em} R}}
\newcommand{\N}{\mbox{\rm I \hspace{-0.8em} N}}

\def\be{\begin{equation}}
\def\ee{\end{equation}}
\def\bea{\begin{eqnarray}}
\def\eea{\end{eqnarray}}
\def\ba{\begin{array}}
\def\ea{\end{array}}
\def\la {\lambda}
\def \part {\partial}
\def \al {\alpha}
\def \de {\delta}


\vskip 5mm

\begin{abstract}
A Hamiltonian pair with arbitrary constants is proposed and thus
a sort of hereditary operators is resulted. All
the corresponding systems of evolution equations
possess local bi-Hamiltonian formulation and a special 
choice of the systems leads to the KdV hierarchy. 
Illustrative examples are given.
\end{abstract}

Bi-Hamiltonian formulation is significant 
for investigating integrable properties
of nonlinear systems of differential equations \cite{Magri} \cite{GelfandD}
\cite{Dickey}.
Many mathematical and physical systems have been found to 
possess such kind of bi-Hamiltonian formulation. There are two important 
problems related to bi-Hamiltonian theory.
The one is which kind of systems can possess bi-Hamiltonian 
formulation and the other one is 
how we construct bi-Hamiltonian formulation for a given system
if it exists.
There has been no complete answer to these two problems so far, 
although a lot of general
analysis for bi-Hamiltonian formulation itself has been made.
However we can make as many observations 
on structures of various bi-Hamiltonian systems as possible,
through which we may eventually find a possible way to the final end.

With such an idea or a motivation
to enhance our understanding of 
bi-Hamiltonian formulation, we would like 
to search for new examples of 
bi-Hamiltonian systems among coupled KdV systems and their higher 
order partners.
There are already some theories which allow us to do that. 
For instance, we can generate soliton 
hierarchies by using decomposable hereditary operators \cite{FuchssteinerF}
or by using perturbation around solutions \cite{MaF1996b}.
In this paper, we would just like to present some new concrete examples 
to satisfy the Magri scheme \cite{Magri} by considering
decomposable hereditary operators. 

Let us choose two specific matrix differential operators:
\be    {J} = \left [ \ba {cccc}  0 & & & \al _0\part  \vspace{2mm}\\
& & \al _0 \part & \al _1 \part  
 \vspace{0.5mm}\\ &\begin{turn}{45}
\vdots
\end{turn} &\begin{turn}{45}
\vdots
\end{turn}  & \begin{turn}{45}
\vdots
\end{turn}   \vspace{3.5mm}
\\
 \al _0 \part &  \al _1  \part &\cdots  & \al _N \part  
 \ea  \right],\ 
   {M} = \left [ \ba {cccc}  0 & & & M_0  \vspace{2mm}\\
& & M _0  & M_1  
 \vspace{0.5mm}\\ &\begin{turn}{45}
\vdots
\end{turn} &\begin{turn}{45}
\vdots
\end{turn}  & \begin{turn}{45}
\vdots
\end{turn}   \vspace{3.5mm}
\\  M _0  &  M _1   &\cdots  & M _N   
 \ea  \right],
  \ee
with 
\be \part =\frac{\part }{\part x},\ 
M_i=c_i\part ^3 +d_i\part +2u_{ix}+4u_i\part ,\ u_i=u_i(x,t),\
0\le i\le N, \ee  
where $\al _i,c_i,d_i,\,0\le i\le N,$ are arbitrary constants, but 
$\al _0\ne 0$ which guarantees the invertibility of $J$. 
It is known \cite{Ma1990a} that 
$  {J}$ and $  {M}$ constitute a pair of Hamiltonian operators with respect to 
the potential vector $u=(u_0,u_1,\cdots ,u_N)^T$, that is to say,
$a J + b M$ is a Hamiltonian operator for any two constants $a $ and $b$,
which may also be proved directly by the Gel'fand-Dorfman algebraic method 
\cite{GelfandD} \cite{Dorfman}.

Now we can generate a hereditary operator $  {\Phi }=  {M}  {J}^{-1}$
(see, say, \cite{FuchssteinerF} for a general proof), since $J$ is invertible.
To express this operator explicitly,
we need to compute the inverse of $  {J}$.  
In view of the specific form of $J$, we can assume 
\be    {J}^{-1} = \left [ \ba {cccc}  \beta _0\part ^{-1} &\beta _1
\part ^{-1} & \cdots & \beta _N\part ^{-1}  \vspace{2mm}\\
\beta _1\part ^{-1} & &
\begin{turn}{45}
\vdots
\end{turn}
 &  \vspace{0.5mm}\\ 
\vdots
&\begin{turn}{45}
\vdots
\end{turn} & &   \vspace{3.5mm}\\
 \beta _N\part ^{-1} &   &  & 0  
 \ea  \right], \ee
where $\beta _i,\ 0\le i\le N$, are constants to be determined.
It is easy to get that
\[ 
J  {J}^{-1}=(J^{-1}J)^T=
 \left [ \ba {cccc}  \al _0\beta _N & & & 0  \vspace{2mm}\\
\al _0\beta _{N-1}+\al _1\beta _N &\al _0\beta _N & &  \vspace{2mm}\\
\vdots &\ddots &\ddots &   \vspace{2mm}\\
 \al _0 \beta _0+\al _1\beta _1+\cdots +\al _N\beta _N & \cdots   & \al _0
\beta _{N-1}+\al _1\beta _N  & 
  \al _0\beta 
 \ea  \right].\]
Therefore $JJ^{-1}=J^{-1}J=I_{N+1},$ where $I_{N+1}$ is an identity matrix 
operator of size $(N+1)\times (N+1)$, leads to an equivalent system
of linear algebraic equations for $\beta _i$, $0\le i\le N$:
\be  \al _0\beta _N =1,\ \al _0\beta _{N-1}+
\al _1\beta _N=0,\ \cdots,\  \al _0 \beta _0+\al _1\beta _1+\cdots +\al _N
\beta _N =0, \ee
which may be written as 
\be A\beta =E_1,\ \textrm{i.e.} \ 
\left [ \ba {cccc}  0& & & \al _0 \vspace{2mm}\\
& & \al _0 &\al _1 
 \vspace{0.5mm}\\ &\begin{turn}{45}
\vdots
\end{turn} &\begin{turn}{45}
\vdots
\end{turn}  &
\vdots
 \vspace{3.5mm}
\\
\al _0&\al _1&\cdots &\al _N \ea
 \right]\left [ \ba {c} \beta_0 
\vspace{2mm}\\ \beta_1 \vspace{2mm}\\ \vdots
 \vspace{2mm}\\ \beta_N \ea
 \right]=\left [ \ba {c}1 \vspace{2mm}\\0 \vspace{2mm}\\ \vdots
 \vspace{2mm}\\0 \ea
 \right].
 \ee
The coefficient matrix $A$ is invertible since $\al _0\ne 0$,
and thus this linear 
system has a unique solution $\beta =A^{-1}E_1$.
Now we can obtain
\be   {\Phi }=
  {M}  {J}^{-1}=
 \left [ \ba {cccc} \beta _N \Phi _0&  & & 0 \vspace{2mm}\\
\beta _{N-1}\Phi _0 +\beta _{N}\Phi _1& \beta _N \Phi _0& & \vspace{2mm}\\
\vdots  &\ddots &\ddots &  \vspace{2mm}\\
\beta _0\Phi _0+\beta _1\Phi _1+\cdots +\beta _N\Phi _N&
\cdots & \beta _{N-1}\Phi _0 +\beta _{N}\Phi _1&\beta _N\Phi _0
 \ea  \right],
\ee 
where 
\be \Phi _i=M_i\part ^{-1}= c_i\part ^2 +d_i+2u_{ix}\part ^{-1}+4u_i, 
\ 0\le i\le N, \ee
and then the conjugate operator of ${\Phi }$ reads as
\be   {\Psi }=  {\Phi }^\dagger = \left [ \ba {cccc} \beta _N\Psi _0 & 
\beta _{N-1}\Psi _0 +\beta _{N}\Psi _1& \cdots & 
\beta _0\Psi _0+\beta _1\Psi _1+\cdots +\beta _N\Psi _N \vspace{2mm}\\
& \ddots &\ddots &\vdots \vspace{2mm}\\
& & \beta _N\Psi _0&\beta _{N-1}\Psi _0 +\beta _{N}\Psi _1\vspace{2mm}\\
0& & & \beta _N\Psi _0\ea \right],
 \ee 
where 
\be \Psi _i=\Phi_i^\dagger =
 c_i\part ^2 +d_i+2u_i+2\part ^{-1}u_{i}\part ,
\ 0\le i\le N. \ee

Because the Lie derivative of $  {\Phi }$ with respect to 
$u_x$ is zero, i.e.  
\[ L_{u_x}\Phi = \left.\frac {\part }{\part \varepsilon }\right|
_{\varepsilon=0}\Phi (u+\varepsilon u_x)
-[I_{N+1}\part ,\Phi ]=0, \]
we have (see, say, \cite{Magri} \cite{Fuchssteiner1979} \cite{Oevel}
\cite{Ma1990b})
\be [K_m,K_n]=
\left.\frac {\part }{\part \varepsilon }\right|
_{\varepsilon=0}\left(K_m(u+\varepsilon K_n)-
K_n(u+\varepsilon K_m)\right)=0,\ K_n:={\Phi }^nu_x=0,\ m,n\ge0.\ee
This also implies that 
a hierarchy of systems of evolution equations $u_t=K_n,\ n\ge 0,$
has infinitely many common commuting symmetries $\{K_m\}_0^{\infty}$.
All systems in the hierarchy have a common recursion operator $\Phi$,
since the operator 
$\Phi $ is hereditary and has a zero Lie derivative with respect to
$u_x$: $L_{u_x}\Phi=0$.
Moreover, due to the specific forms of $\Phi _i,\,0\le i\le N$, 
they are all local, although the recursion operator $\Phi$
is integro-differential. A mathematical induction process may easily 
verify this statement on locality.

In what follows, we want to show local bi-Hamiltonian formulation for 
all systems but the first one in the hierarchy (note that 
sometimes systems of soliton equations just have one local Hamiltonian 
formulation in bi-Hamiltonian formulation, such examples can be 
the modified KdV equations and O(3) chiral field equations
\cite{BordagY}). First of all,
we observe the second system
\[ u_t=K_1=  {\Phi }u_x=  {M}  {J}^{-1}u_x.\]
The vector field $J^{-1}u_x$ can be computed as follows
\bea   {J}^{-1}u_x&=&
\left [ \ba {cccc}  \beta _0\part ^{-1} &\beta _2
\part ^{-1} & \cdots & \beta _N\part ^{-1}  \vspace{2mm}\\
\beta _1\part ^{-1} & & 
\begin{turn}{45}
\vdots \end{turn} &  \vspace{0.5mm}\\
\vdots 
&\begin{turn}{45} \vdots \end{turn} & &   \vspace{3.5mm}\\
 \beta _N\part ^{-1} &   &  & 0  
 \ea  \right]\left [ \ba {c}u_{0x} \vspace{2mm}\\
u_{1x} \vspace{2mm}\\ \vdots \vspace{2mm}\\u_{Nx}\ea  \right]
\nonumber \\ 
&=&\left [ \ba {c}\beta _0u_0+\beta _1u_1+\cdots +\beta _Nu_N \vspace{2mm}\\
\beta _1u_0+\beta _2u_1+\cdots +\beta _Nu_{N-1}
 \vspace{2mm}\\ \vdots \vspace{2mm}\\ \beta _{N-1}u_0+\beta _N u_1
\vspace{2mm}\\ \beta _N u_0 \ea  \right]\nonumber \\
& = &\beta _N
\left [ \ba {c}u_N  \vspace{2mm}\\ u_{N-1} \vspace{2mm}\\ 
u_1 \vspace{2mm}\\ \vdots \vspace{2mm}\\ u_0 \ea  \right]+\beta _{N-1}
\left [ \ba {c}u_{N-1}  \vspace{2mm}\\ u_{N-2} \vspace{2mm}\\ 
\vdots \vspace{2mm}\\ u_0\vspace{2mm}\\ 0 \ea  \right]
+\cdots +\beta_0
\left [ \ba {c}u_{0}  \vspace{2mm}\\ 0 \vspace{2mm}\\ 
\vdots \vspace{2mm}\\ 0\vspace{2mm}\\ 0 \ea  \right]\nonumber \\ 
&{\buildrel {\textrm{def}}\over  =}& 
\beta _NX_0+\beta _{N-1}X_1+\cdots +\beta _0X_N.
\eea 
Evidently we can find or directly prove that
\be  f_0:=  {J}^{-1}u_x=\frac {\delta \tilde{H}_0}{\delta u},\ee
\be \tilde{H}_0=\int H_0\,dx,\  H_0=
\int_0^1<f_{0}(\la u),u >d\lambda =
\frac12 \sum_{l=0}^N\beta _l\sum_{i+j=l}u_iu_j,\label{H0}\ee
where $<\cdot,\cdot >$ denotes the standard inner product of $\R ^{N+1}$.
This means that $f_0$ is gradient. Now
we check whether or not the vector field $  {\Psi }f_0$
is a gradient field, which is required in the Magri scheme \cite{Magri}.
A direct computation can give  
\bea &&  f_{1m}:=  {\Psi }X_{m}= 
\frac {\delta \tilde{H}_{1m}}{\delta u},\ \tilde{H}_{1m}= 
\int _{-\infty}^\infty H_{1m}\,dx, \nonumber \\&&
 H_{1m}=\int_0^1<f_{1m}(\la u),u >d\lambda 
\nonumber \\&& = \sum_{l=m}^N\beta _l\sum_{i+j+k=l-m}
\left[ \frac12 (c_iu_{j}u_{kxx}+d_iu_ju_k)+u_iu_ju_k\right], \ 0\le m\le N. 
\nonumber \eea
These equalities yield
\be f_1:=  {\Psi }f_0=\frac {\delta \tilde{H}_{1}}{\delta u},  \ee
where the Hamiltonian functional $\tilde{H}_1$ is determined by
\be \tilde{H}_{1}=
\int H_1\,dx,\ 
H_1=\sum_{m=0}^N\beta _{N-m}\sum_{l=m}^N\beta _l\sum_{
i+j+k=l-m}
\left[ \frac12 (c_iu_{j}u_{kxx}+d_iu_ju_k)+u_iu_ju_k\right].  \label{H1}\ee 
Therefore the system $u_t=K_1=\Phi u_x$ has local bi-Hamiltonian
formulation 
\be u_t=K_1=\Phi u_x= J\frac {\delta \tilde{H}_{1}}{\delta u}
=M\frac{\delta \tilde{H}_{0}}{\delta u}, \ee
where $\tilde{H}_0$ and $\tilde{H}_1$ 
are defined by (\ref{H0}) and (\ref{H1}), respectively.

Secondly, we want to expose bi-Hamiltonian formulation for the other systems
$u_t=K_n,\ n\ge 2$. Note that $\Phi =\Psi ^\dagger$ is hereditary, and 
that $f_0$ and $\Psi f_0$ are already gradient.
According to the Magri scheme \cite{Magri} \cite{FuchssteinerF},
all vector fields $  {\Psi }^nf_0$, $n\ge 0$,
are gradient fields, namely there exists a hierarchy of functionals 
$\tilde{H}_n$, $n\ge 0$, such that
\be f_n:=  {\Psi }^nf_0= \frac {\delta \tilde{H}_{n}}{\delta u},\ n\ge 0.
\ee
In fact, the Hamiltonian functionals $\tilde{H}_n, \ n\ge 0,$ must be equal to
\be \tilde{H}_{n}=
\int  H_n\,dx,\ 
H_{n}=\int_0^1<f_{n}(\la u),u >d\lambda , \ n\ge 0,\ee
and they are all in involution with respect to 
either Poisson bracket:
\bea && \{\tilde{H}_m,\tilde{H}_n\}_J:=
\int \frac{\delta \tilde{H}_m}{\delta u}J\frac{\delta 
\tilde{H}_n}{\delta u}\,dx=0,\ m,n\ge 0,\\ && 
 \{\tilde{H}_m,\tilde{H}_n\}_M:=
\int \frac{\delta \tilde{H}_m}{\delta u}M\frac{\delta 
\tilde{H}_n}{\delta u}\,dx
=0,\ m,n\ge 0. \eea
This way all vector fields $K_n,\ n\ge 1,$ can be written in two ways
as 
\bea && K_n={\Phi }^nu_x={\Phi }^nJf_0=
 {J} \Psi ^n f_0=J\frac {\delta \tilde{H}_{n}}{\delta u},\ n\ge 1,
\nonumber \\&& 
   K_n={\Phi }^nu_x=(J\Psi )\Psi ^{n-1}f_0= {M}   {\Psi }^{n-1}
f_0=   {M}\frac {\delta \tilde{H}_{n-1}}{\delta u},\ n\ge 1, \nonumber \eea
which provide local bi-Hamiltonian formulation  
\be 
u_t=K_n=  {\Phi }^nu_x=
 {J} \frac {\delta \tilde{H}_{n}}{\delta u}
=   {M}\frac {\delta \tilde{H}_{n-1}}{\delta u},\ n\ge 1,
\ee
for all systems
$u_t=K_n,\,n\ge 1$. 
It follows that the systems $u_t=K_n,\, n\ge 0,$ have common 
commuting symmetries and conserved quantities, which justifies that
they constitute a typical soliton hierarchy.  

Note that the coefficients appearing in our construction are all arbitrary
except the requirement of $\al _0\ne 0$. Thus the resulting systems 
may contain many interesting systems. A special choice of 
\[ \al _0=c_0=1,\ \al _i=c_i=0,\ 1\le i\le N,\ d_i=0,\ 0\le i\le N,\] 
leads to the KdV hierarchy under the reduction $u_i=0,\, 1\le i\le N,$
and thus the above resulting systems are called coupled KdV systems.

Let us now show some examples.
Let $N=0$ and $\al _0=1$. At this moment, we have
\[   {J} =\part , \   {M}=c_0\part ^3+d_0\part +2u_{0x}+4u_0\part 
,\ 
  {\Phi }=c_0\part ^2+d_0 +2u_{0x}\part ^{-1}+4u_0 .\]  
When $c_0\ne 0,\, d_0=0$, the corresponding hierarchy is the KdV hierarchy.
When $c_0=  d_0=0$, the corresponding hierarchy is
a hierarchy of quasi-linear partial differential equations, of which 
the first two nonlinear equations are 
\[ u_t=  {\Phi }u_x=6u_0u_{0x},\ u_t=  {\Phi }^2u_x= 30u_0^2u_{0x}.\]
All the vector fields and all 
the Hamiltonian functionals in this hierarchy are of special 
form $cu_0^mu_{0x}$ and 
$cu_0^m$, where $c$ is a constant and $m\in \N $, respectively.

Let $N=1$. The corresponding Hamiltonian pair and hereditary operator
become
\bea &&    {J}=\left [ \ba {cc}0& \al _0 \part \vspace{2mm}\\ 
\al _0\part & \al _1\part \ea \right],\ 
   {M}=\left [ \ba {cc}0& c_0\part ^3 +d_0\part +u_{0x}+4u_0 \part
 \vspace{2mm}\\ 
  c_0\part ^3 +d_0\part +u_{0x}+4u_0 \part & 
 c_1\part ^3 +d_1\part +u_{1x}+4u_1 \part \ea \right],\nonumber \\ &&
   {\Phi }=  {M}  {J}^{-1}=
  {M}\left [ \ba {cc}-\frac {\al _1}{\al _0^2}\part ^{-1}
& \frac 1{\al _0} \part ^{-1} \vspace{2mm}\\ 
\frac 1 {\al _0}\part ^{-1} &0 \ea \right]
=\left [ \ba {cc}
\frac 1 {\al _0}\Phi _0&0 \vspace{2mm}\\ 
- \frac {\al _1}{\al _0^2}\Phi_0 +
\frac 1 {\al _0}\Phi _1&
\frac 1 {\al _0}\Phi _0
\ea \right],\nonumber 
\eea 
where $\Phi_0=c_0\part ^2 +d_0 +2u_{0x}\part ^{-1}+4u_0 $ and $
\Phi _1= c_1\part ^2 +d_1 +2u_{1x}\part ^{-1}+4u_1 $.
The first nonlinear system is the following
\[\ba {l} 
u_t=  {\Phi }u_x  =  {J}\frac {\delta \tilde{H}_1}{\delta u}=  {M}
\frac {\delta \tilde{H}_0}{\delta u}\vspace{2mm}\\
=\left [ \ba {c} \frac 1 {\al _0}
(c_0u_{0xxx}+d_0u_{0x}+6u_0u_{0x}) \vspace{2mm}\\
 - \frac {\al _1}{\al _0^2}( c_0u_{0xxx}+d_0u_{0x}+6u_0u_{0x})
+ \frac 1 {\al _0}[(c_0u_1+c_1u_0)_{xxx}+(d_0u_1+d_1u_0)_x+6(u_0u_1)_x]
\ea \right], \ea \]
where the Hamiltonian functionals read as
\bea &&
\tilde{H}_0=
\int H_0\,dx,\ H_0=  \frac 1 {\al _0}u_0u_1-\frac {\al _1}{2\al _0^2}
u_0^2, \nonumber \\ && 
\tilde{H}_1=\int H_1\,dx,\ H_1=
\frac 1 {\al _0^2}(\frac {c_1}2-\frac {c_0\al _1}{\al _0})u_0u_{0xx}
+\frac 1 {\al _0^2}(\frac {d _1}2-\frac {d _0\al _1}{\al _0})u_0^2-
\frac {2\al _1}{\al _0^3}u_0^3
 \nonumber \\ && 
 \qquad \qquad\qquad\quad  
+\frac 1 {\al _0^2}\left[\frac {c_0}2(u_0u_{1xx}
+u_{0xx}u_1)+d_0u_0u_1+3u_0^2u_1\right]
 . \nonumber \eea

For a general case of $N$, if we choose 
\[\left \{ \ba {l} \al _0=1,\ \al _i=0,\ 1\le i\le N,\vspace{2mm}\\
 c_0=1,\ c_i=0,\ 0\le i\le N, \vspace{2mm}\\ d_i=0,\ 
0\le i\le N, \ea \right. \] 
then the resulting systems are exactly the perturbation systems 
of the KdV hierarchy introduced in Ref. \cite{MaF1996a} through perturbation 
around solutions of the KdV equation.

It should be realized that all first nonlinear systems ($u_t=K_1=\Phi u_x$)
belong to a more general class of integrable coupled KdV systems,
which was introduced by G\"urses and Karasu 
in \cite{GursesK-JMP1998}, motivated by the Jordan KdV systems
in \cite{Svinolupov}.
Moreover the principle part of our coupled KdV systems, i.e.
the systems with $d_i=0,\,0\le i\le N$, 
belong to a symmetric subclass in the non-degenerate case 
in \cite{GursesK-JMP1998}.
This may be seen by observing the coefficients 
\be b^i_j=\sum_{k=0}^N\beta _{N-i+j+k}c_k, \ a_{jk}^i=2c_{jk}^i
=\frac 23 s_{jk}^i= 4\beta _{N-i+j+k}, \ 0\le i,j,k\le N, 
\label{coefficientsofbac_{ijk}}\ee
where $\beta _i=0,\ i<0\ \rm{or}\ i>N$, are accepted,
after our recursion operators 
and our coupled systems in the case of 
$d_i=0,\,0\le i\le N$,
are rewritten as follows
\bea &&
\Phi (u)=(R^i_j)_{(N+1)\times (N+1)},\ 
R_j^i= b_j^i\part ^2+\sum_{k=0}^N(a_{jk}^iu_k+c_{jk}^iu_{kx}\part ^{-1}),
\label{typerecursionoperator}\\ &&
u_{it}=\Phi (u)u_x=\sum_{k=0}^Nb^i_ku_{kxxx}+\sum_{k,j=0}^Ns_{jk}^iu_ju_{kx},
\ 0\le i\le N.  \label{typecoupledKdVsystem}\eea
Actually the coefficients defined by (\ref{coefficientsofbac_{ijk}})
satisfy the relations 
\[ \sum_{k=0}^Nb^k_ls_{jk}^i=\sum_{k=0}^N b_k^is_{jl}^k,\ 
\sum_{k=0}^N s_{jk}^is_{lm}^k
=\sum_{k=0}^N s_{lk}^is_{jm}^k,\ 0\le i,j,l,m\le N,
 \]
which guarantees \cite{GursesK-JMP1998} that the operators 
defined by (\ref{typerecursionoperator})
with the coefficients 
$a_{jk}^i=2c_{jk}^i
=\frac 23 s_{jk}^i$
are recursion operators
for the systems determined by (\ref{typecoupledKdVsystem}) in the symmetric 
case of $s_{jk}^i=s_{kj}^i$.

The other nonlinear systems in each hierarchy determined by 
a hereditary operator $\Phi $ may contain much 
higher order derivatives of $u$ with respect to $x$, but they still have a 
recursion operator and even bi-Hamiltonian formulation. 
By taking a scaling transformation $t\to at,\ x\to bx,\ u\to cu$,
more concrete examples of integrable coupled KdV systems 
\cite{GursesK-JMP1998} can be obtained from our systems.

Compared with the well-known coupled KdV systems (for example, see 
\cite{BoitiCP,AntonowiczF,Ma1993b}),
the above systems are not really coupled because of the first
separated component.
The Lax pairs or the spectral problems associated with our systems
have not been found yet. If they are found,
master symmetries of the systems can also be presented like
ones of the well-known coupled KdV systems in \cite{Ma1993b}. 

Using an idea of extension in Ref. \cite{GhoshC}, we may 
obtain much more general Hamiltonian pairs starting from the above one. 
Also we may have other choices, say,
\[\left [\ba {cc } r_x+2r\part & s\part \vspace{2mm} \\ 
s_x+s\part & 0  \ea \right]+
\left [\ba {cc } c_1 & c_2 \vspace{2mm} \\ 
c_2 & c_3  \ea \right]\part
 +\left [\ba {cc } 0& c_4 \vspace{2mm} \\ 
-c_4 & 0  \ea \right]\part ^2, \ u=\left [\ba {c }
 r \vspace{2mm} \\ s  \ea \right],\ c_i=\textrm{consts.},\ 1\le i\le 4, 
\]
in Ref. \cite{Ma1993c}, or 
start from more general Hamiltonian operator structures
to construct new integrable systems having bi-Hamiltonian formulation,
but we need more techniques in manipulation. 

\vskip 3mm
\noindent{\bf Acknowledgment:} The author
would like to thank the
City University of Hong Kong and the Hong Kong Research Grants Council 
for financial support.



\small 
\baselineskip 13pt
\begin{thebibliography}{99}
\bibitem{Magri}Magri F 1978 {\it J. Math. Phys.} {\bf 19} 1156--1162,
1980 in: {\it Lectures Notes in Physics} Vol. 120 (Berlin:
Springer-Verlag) 
pp233--263
\bibitem{GelfandD}Gel'fand I M and Dorfman I Y 1979 
{\it Funct. Anal. Appl.} {\bf 13} 248--262, 1981 {\bf 15} 173--187
\bibitem{Dickey} Dickey L A 1991 
{\it Soliton Equations and Hamiltonian Systems}
(Singapore: World Scientific)
\bibitem{FuchssteinerF} Fuchssteiner B and Fokas A S 1981 {\it Physica D}
{\bf 4} 47--66
\bibitem{MaF1996b} Ma W X and Fuchssteiner B 1996
{\it Chaos, Solitons $\&$ Fractals} {\bf 7} 1227--1250
\bibitem{Ma1990a} Ma W X 1990 
{\it Acta Math. Appl.
Sinica} {\bf 13} 484--496
\bibitem{Dorfman} Dorfman I Y 1993 {\it Dirac Structures and Integrability
of Nonlinear Evolution Equations} (Wiley, New York)
\bibitem{Fuchssteiner1979} Fuchssteiner B 1979
{\it Nonlinear Anal. Theor. Meth. Appl.} {\bf 3} 849--862
\bibitem{Oevel} Oevel W 1987 in: {\it Nonlinear Evolution Equations, Solitons
and the Inverse Scattering Transform}, eds. Ablowitz M, Fuchssteiner B and
Kruskal M (Singapore: World Scientific) pp108--124
\bibitem{Ma1990b} Ma W X 1990 {\it J. Phys. A: Math. Gen.} {\bf 13} 2707--2716 
\bibitem{BordagY}
Bordag L A and Yanovski A B 1996 {\it J. Phys. A: Math. Gen.} {\bf 29}
5575--5590
\bibitem{MaF1996a} Ma W X and Fuchssteiner B 1996
{\it Phys. Lett. A} {\bf 213} 49--55
\bibitem{GursesK-JMP1998} G\"urses M and Karasu A 1998
  {\it J. Math. Phys.} {\bf 39} 2103--2111
\bibitem{Svinolupov}Svinolupov S I 1991 {\it Theor. Math. Phys.} {\bf 87}
611--620
\bibitem{BoitiCP} Boiti M, Caudrey P J and Pempinelli F 1984 
{\it Nuovo Cimento B} {\bf 83} 71--87
\bibitem{AntonowiczF} Antonowicz M and Fordy A P 1987 {\it Physica D} 
{\bf 28} 345--357
\bibitem{Ma1993b} Ma W X 1993 {\it J. Phys. A: Math. Gen.} {\bf 26}
2573--2582
\bibitem{GhoshC} Ghosh C and Chowdhury A R 1994
{\it J. Phys. Soc. Jpn.} {\bf 63} 3911--3913
\bibitem{Ma1993c} Ma W X 1993 {\it Applied  Mathematics -- A Journal 
of Chinese Universities} {\bf 8} 28--35
\end{thebibliography}
\end{document}